\documentclass{PoS}

\title{(Lattice) Propagators and
Extraction of Spectral
Densities}

\ShortTitle{Lattice Spectral Densities}

\author{\speaker{David Dudal}\thanks{I wish to thank J.~Greensite and the organizers for the opportunity to speak at this meeting.}\\
        Ghent University, Department of Physics and Astronomy, Krijgslaan 281-S9, B-9000 Gent, Belgium\\
        E-mail: \email{david.dudal@ugent.be}}

\author{Orlando Oliveira\\
        Centro de F\'{i}sica Computacional, Departamento de F\'{i}sica, Universidade de Coimbra, 3004-516 Coimbra, Portugal\\
        E-mail: \email{orlando@teor.fis.uc.pt}}

\author{Paulo J.~Silva\\
        Centro de F\'{i}sica Computacional, Departamento de F\'{i}sica, Universidade de Coimbra, 3004-516 Coimbra, Portugal\\
        E-mail: \email{psilva@teor.fis.uc.pt}}

\abstract{In this proceeding, we explain a few steps for an alternative extraction of the spectral density of a two-point function (propagator) based on a discrete set of data points. We present a so-called Tikhonov regularization of this particular inverse problem. We test it on 2 cases: lattice $0^{++}$ glueball data and mock gluon data.}

\FullConference{Xth Quark Confinement and the Hadron Spectrum,\\
		October 8-12, 2012\\
		TUM Campus Garching, Munich, Germany}

\begin{document}

\section{Introduction}
Throughout this proceeding ---see also \cite{Oliveira:2012eu}--- we shall be interested in a generic Euclidean space two-point function denoted by $D(p^2)$. We leave it open whether this corresponds to the propagation of elementary particles or bound state degrees of freedom, corresponding to a certain Lagrangian of a set of fundamental fields. In any case, this $D(p^2)$ should obey a very specific integral representation, given by
\begin{equation}\label{1}
    D(p^2)=\int_0^\infty d\mu\frac{\rho(\mu)}{\mu+p^2},
\end{equation}
at least when the considered particle is one that belongs to the asymptotic observable spectrum.

This integral is the (Euclidean version) of the K\"all\'{e}n-Lehmann representation, with $\rho(\mu)$ the spectral density. In a more mathematics based language, Eq.~(\ref{1}) corresponds to a Stieltjes integral transform \cite{widder}. Using Cauchy's theorem, it is quickly derived that $\rho(t)$ is directly proportional to the discontinuity of $D(t)$ in the complex plane along the negative real axis. Through the optical theorem, this has the important consequence that $\rho(\mu)$ must be positive as it corresponds to a physical probability. The spectral density $\rho(\mu)$ contains crucial physical information. For example, a massive particle will reflect in a $\delta(\mu-\mu_0)$-function peak in $\rho(\mu)$, finite peaks can be related to instable particles etc. Spectral densities are also useful to study transport properties of e.g.~the quark gluon plasma and so on.

The expression (\ref{1}) defines the unique analytic continuation to the complex squared momentum plane of the Euclidean propagator that is usually\footnote{In particular when using lattice Monte Carlo simulations.} only probed for $p^2\geq 0$. However, the behaviour of Green functions in terms of complex momenta is not only of importance for the study of the physical spectrum of quantum field theories, knowledge of these is also of paramount importance to study the bound state equations in terms of their constituents. Let us only think about unphysical gluons and quarks that are however bound together into colorless states.  These can, for example, be studied using moment techniques \cite{nogeenboek,Dudal:2010cd} or the Bethe-Salpeter equations \cite{Maris:2003vk}.  Precise spectral knowledge of the input gluons and/or quarks is required in these instances.

We hope that by now we have briefly motivated why it is important to learn about the spectral properties of Green functions, in particular of propagators. It is however a very cumbersome task to directly convert (accessible) knowledge of $D(p^2)$ into an estimate for $\rho(\mu)$. A beautiful result of Widder (1941) says
\begin{equation}
\rho(t)=\lim_{n\to +\infty} (-1)^{n+1} \frac{1}{(n!)^2} \partial_t^{n}\left[t^{2n+1}\partial_t^{n+1}D(t)\right]\,,\qquad t\geq0
\end{equation}
but this formula is in most cases impractical: taking the $\infty$th derivative is a numerical beast. In some cases stability can be reached for large $n$ using appropriate numerical derivation and some analytical cases can be worked out exactly, see f.i.~\cite{Dudal:2010wn}.

The main goal is, in principle, when $D(p^2\geq0)$ is sufficiently well known via nonperturbative lattice data, to construct an trustworthy approximation for $\rho(\mu)$, i.e.~we wish to obtain at least some information on the analytical structure of Green functions.  A powerful method has been advocated in \cite{Asakawa:2000tr}, based on the \emph{Maximum Entropy Method} (MEM) from inverse problem analysis.  Although powerful, MEM is also a quite demanding tool from the calculational viewpoint, and in its original formulation it assumes a positive spectral density. Therefore, we will try to develop a somewhat less demanding tool that simultaneously could also be used to get e.g.~spectral information on unphysical particles.

\section{Tikhonov-Morozov regularization}
Before turning to physical applications, we introduce in brief the necessary language we shall use. Let us first try to understand better why it is so hard to invert the K\"all\'{e}n-Lehmann integral. We notice that, upon taking a Fourier transform of Eq.~(\ref{1}), we find
\begin{equation}\label{KL3}
f(t)\equiv  \frac{1}{2\pi}\int_{-\infty}^{\infty} d p e^{-i pt} D(p^2)=\frac{1}{2}\int_{0}^{\infty}d\mu\frac{\rho(\mu)}{\sqrt{\mu}}e^{-t\sqrt{\mu}}=\int_{0}^{\infty}d y\rho(y^2)e^{-ty}\equiv\int_{0}^{\infty}d y\hat{\rho}(y)e^{-ty}.
\end{equation}
With the notation $\mathcal{L}f(t)=\int ds e^{-st}f(s)=$ for the (self-adjoint) Laplace transform, we thence observe that
\begin{eqnarray}\label{sub2}
  \mathcal{G}=\mathcal{L}^2\hat\rho=\mathcal{L}\mathcal{L}^\ast \hat\rho,
\end{eqnarray}
meaning that the K\"all\'{e}n-Lehmann representation is nothing else than a double Laplace transform. From this, we can easily appreciate the intrinsic problem in inverting it: it is indeed well-known that $\mathcal{L}^{-1}$ is a typical \emph{ill-posed} problem. The obstacle to simply implement $\mathcal{L}^{-1}$ is due to the exponential dampening: a tiny variation in $\mathcal{L}f$ ($\sim$ propagator) will ask for massive changes in $f$ ($\sim$ spectral density). In practice, as the data for the propagator will always be subject to uncertainties (errors), there will unavoidably appear large uncertainties in the outcome, viz.~$\rho(\mu)$.

In order to give a meaningful version of the inverse K\"all\'{e}n-Lehmann representation, consider a generic ill-posed problem
\begin{equation}\label{tik1}
    y=\mathcal{K}x\,,\quad ||y-y^\delta||\leq \delta,
\end{equation}
with $y^\delta$ a set of data we have for the quantity $y$, polluted with ``noise'' (errors) $\delta$. We now want a controllable solution for $x$: if we get better and better data, we should come closer and closer to the exact solution. A direct inversion is useless, thus we need to \emph{regularize} the system \cite{boek}.

The MEM is one example of obtaining a regularized system. Here, we focus on a (simple)  \emph{Tikhonov regularization}. We will search for a solution $x^\delta$ such that
\begin{equation}\label{tik2}
    \mathcal{J}_\lambda=||\mathcal{K}x-y||+\lambda||x||^2
\end{equation}
is minimal, $\lambda>0$ is a regularization parameter. The meaning of this is intuitively clear: we search for a solution ``sufficiently close'' to the real one in norm, since for $\lambda=0$ we are back at the original (but ill-posed) problem. This already signals that we should not take $\lambda$ too small, at the risk of ending up again close to bad behaviour. Contrarily, for $\lambda$ too large we are drifting too far off the original problem.

One can now show \cite{boek} that $x^\lambda$ is the (unique) solution of the so-called \emph{normal equation}:
\begin{equation}\label{tik3}
    \lambda x^\lambda +\mathcal{K}^\ast \mathcal{K} x^\lambda =\mathcal{K}^\ast y
\end{equation}
Here, we immediately notice that this latter equation is well-posed because the operator $\mathcal{K}^\ast \mathcal{K}+\lambda$ is strictly positive and as such invertible. One can also understand the importance of the strictly positive parameter $\lambda$ from a matrix point of view. If the matrix representation of the start operator $\mathcal{K}$ would be considered, a singular value decomposition would reveal the occurrence of near-to-zero singular values, obstructing the inversion. The presence of $\lambda>0$ does however acts as a screening filter on these very small singular values.

Concerning the choice of $\lambda$, we employ an a posteriori fixing by making use of the solution $x^\lambda$. A controllable way is the \emph{Morozov discrepancy principle}: choose that $\lambda$ with
\begin{equation}\label{tik4}
    ||\mathcal{K}x^\lambda-y^\delta||=\delta
\end{equation}
A unique solution $x^{\lambda,\delta}$ exists as can be proven. This again looks quite reasonable: if the noise on the input data vanishes, $\delta\to0$, the ``noise'' on the approximate equation will also vanish. We basically search for ``output'' of similar quality as the ``input''. The discrepancy principle also avoids selecting a too small $\lambda$, which would drive us back dangerously close to the ill-posed case.

We end with a few small observations.
\begin{itemize}
\item The assumption made is that the integral equation (\ref{1}) has a solution. This is of course not guaranteed, we shall come back to this issue later.
\item The equations to be solved become \emph{linear}, so the computational effort is well under control. It also allows to use standard tools to approximate the integrals etc. We shall elaborate on this in a paper in preparation \cite{nieuw}.
\item There are no a priori assumptions made on the sign of solution, i.e.~$\rho(\mu)$ can thus be non-positive.
\end{itemize}

\section{A first application: the lattice scalar glueball}
We can now test our proposed method on a physical example. We consider pure SU(3) gauge theory. The physical quantities are believed to be colorless gauge invariant gluon bound states, better known as \emph{glueballs}. The simplest (lightest) one should be the scalar state, with $J^{PC}$ quantum numbers $0^{++}$. We constructed the lattice equivalent (up to $a$-corrections, with $a$ the lattice spacing) of the corresponding gauge invariant continuum operator $F_{\mu\nu}^2$. We computed its correlator numerically using Monte-Carlo simulations. We therefore generated $\sim 900$ configurations in Coimbra \& Ghent at $\beta=6.2$, $V=64^4$ with $a\approx 0.07261 \textrm{fm}$, $La\approx 4.65 \textrm{fm}$. We are eventually interested in its spectral density where
\begin{equation}
    D(p^2)=\left\langle F_{\mu\nu}^2 F_{\alpha\beta}^2\right\rangle_p=\int_0^{\infty}\frac{\rho(\mu)}{\mu+p^2}d\mu
\end{equation}
A peak in $\rho(\mu)$ can be used to get an estimate of the mass.

We chose this particular example as it can be compared with alternative and well-established mass estimates for the pure glueball states, see e.g.~\cite{Chen:2005mg}.

In $4d$ quantum field theory, there is an additional complication that we need to take into account: ultraviolet divergences need to be tamed by considering a suitably subtracted correlation function. We refer to \cite{nieuw} for the procedure we followed here, we suffice by pointing out that a power counting analysis learns that the used $0^{++}$ correlator asks for 3 subtractions. The bare data is shown in Fig.~1.
\begin{figure}[t]
   \centering
       \includegraphics[width=7cm]{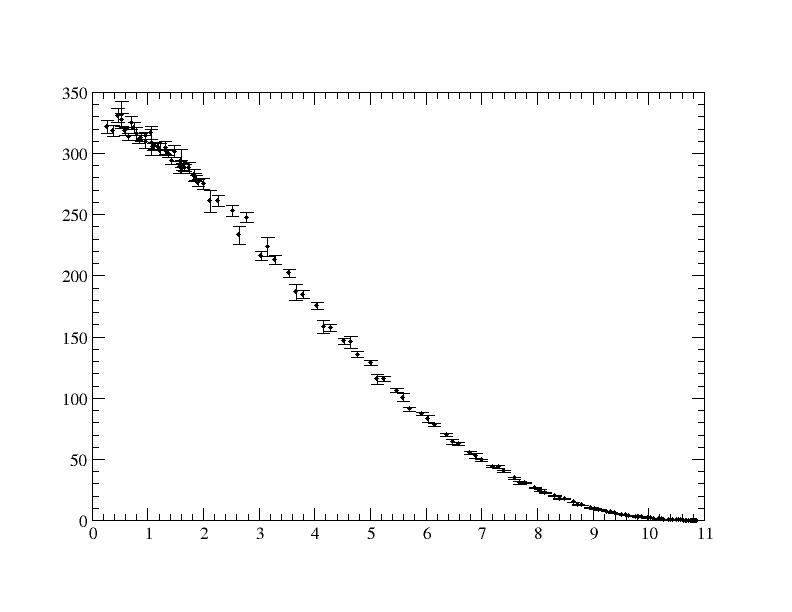}
               \caption{The (bare) scalar glueball correlator.}
\end{figure}
Let us call $\mathcal{G}(p^2)$ the subtracted propagator. Thus we are interested in
\begin{equation}
\mathcal{G}=\mathcal{L}^2\hat\rho.
\end{equation}
The normal equation to be solved reads
\begin{equation}\label{tikons1}
    \mathcal{L}^4 \hat\rho +\lambda\hat\rho=\mathcal{L}^2\mathcal{G}^{\delta},
\end{equation}
or explicitly, this amounts to consider
\begin{equation}\label{tikons2}
    \int_0^\infty dt\hat\rho(t)\frac{\ln\frac{z}{t}}{z-t}+\lambda\hat\rho(z)=\int_0^\infty dt \frac{\mathcal{G}(t)}{t+z}.
\end{equation}
The notation $\hat\rho$ refers to the subtracted spectral density. The r.h.s.~integral (over the data) can be handled with spline interpolation, for the l.h.s.~a stable approach will rely on a Gauss-Chebyshev quadrature.
\begin{figure}[h]
   \centering
       \includegraphics[width=7cm]{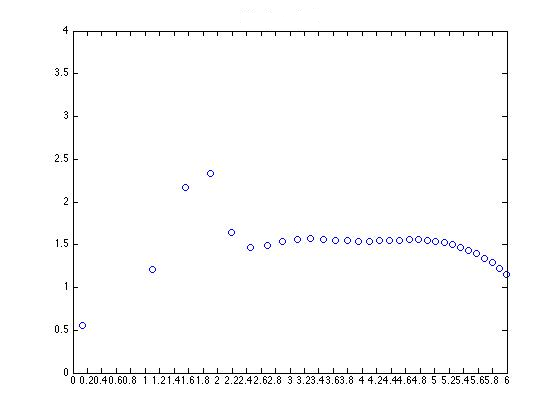}
               \caption{The (subtracted) scalar spectral density. This result is preliminarily and using only part of the data and a simpler numerical integration routine than the one designated in the main text. We also  chose by hand a reasonable (neither too small or too big) value for $\lambda$ for a first trial. A more complete and improved version is currently being prepared \cite{nieuw}.}
\end{figure}
Although the shown (subtracted) spectral density is only a first (incomplete) attempt, we nonetheless observe a clear signal of a peak emerging around $\sim 1.8~\textrm{GeV}$, which is pretty close to what was reported in \cite{Chen:2005mg}, $\sim 1.71~\textrm{GeV}$.

\section{A second application: the mock Landau gauge gluon}
In the case that one desires to tackle the QCD bound state problem with analytical continuum methods, one is forced to abandon gauge invariance in favor of a gauge fixed setting. A widely scrutinized gauge fixing is the Landau choice where the gauge field is made transverse. As trustworthy nonperturbative data is wished also for those gauge fixed quantities, the Landau gauge is a good choice as it can be accessed through lattice sampling as well. In recent years, see \cite{Cucchieri:2007md} for a very short list of references, evidence accumulated in favour of a particular behaviour of the gluon propagator for $p^2\geq 0$.

Different analytical nonperturbative approaches have been considered. One of them is the effective action that takes into account the Gribov gauge copy ambiguity \cite{Dudal:2008sp}, which allows for loop computations whereas nonperturbative effects are taken into account by means of vacuum condensates. This approach also allows to probe the Green functions in the complex squared momentum plane and provides for (tree level) gluon propagators with complex conjugate poles \cite{Oliveira:2012eh}, something which can be interpreted as a manifestation of gluon confinement, see for example also \cite{Bhagwat:2002tx} for a similar quark observation. Attempts to interpret these $cc$ gluon poles in terms of them generating physical bound states can be found in e.g.~\cite{Dudal:2010cd,Capri:2012hh,Windisch:2012sz}.

Recently, also a first attempt in getting information in the complex plane for the gluon propagator was done in \cite{Strauss:2012dg} by solving numerically the functional quantum equations of motion (Dyson-Schwinger equations) for complex $p^2$, reporting no trace of $cc$ poles. As we explained, getting sensible results for spectral densities is a highly complicated task, so it would be worthwhile to test the estimations of \cite{Strauss:2012dg}. Since we have in the meantime developed the machinery to access also non-positive spectral densities with lattice data, we are in principle armed to verify the outcomes of \cite{Strauss:2012dg}, under the assumption that the gluon propagator only displays a cut along the negative real axis, as it was put forward in\footnote{Notice that lattice data alone do not allow to ``prove'' where the cuts would be located. One can however \emph{assume} a specific location and then construct the corresponding spectral density.} \cite{Strauss:2012dg}.

As a kind a warming up exercise for this proceeding, we opted to present an interesting illustration of how delicate things can get when using data. Rather than using already the lattice data itself, we consider the Refined Gribov-Zwanziger fit,
\begin{equation}\label{fot}
    D(p^2)=\frac{p^2+M^2}{p^4+(M^2+m^2)p^2+\lambda^4},
\end{equation}
with the numbers taken from the paper of Dudal, Oliveira and Rodriguez-Quintero \cite{Oliveira:2012eh} and we ``generated'' 40 data points from this\footnote{We omitted a global rescaling factor.}. Since the propagator (\ref{fot}) displays $cc$ poles, it simply cannot have an integral representation of the form (\ref{1}). Nonetheless, using a limited set of data points lying on the curve described by the function (\ref{fot}) does allow to construct a very good approximation assuming that the representation (\ref{1}) does exist after all, see Fig.~3. The corresponding spectral density ---which is non-positive--- is shown thereafter in Fig.~4. Here, we set by hand $\lambda=1/16$.
 \begin{figure}[h]
   \centering
       \includegraphics[width=7cm]{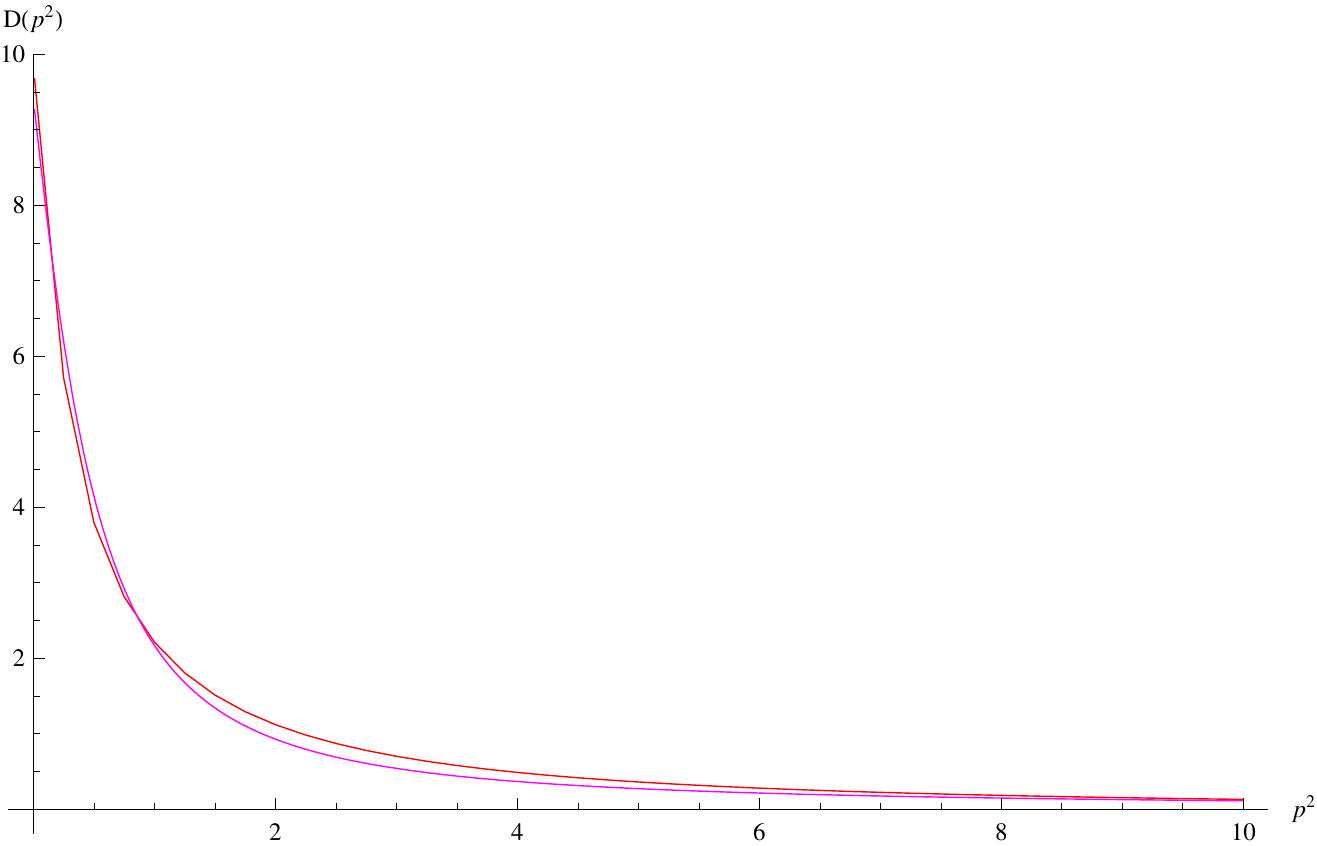}
               \caption{Gluon propagator (magenta) (4.1) and its (in principle non-existent) integral representation (1.1) (red).}
\end{figure}
\begin{figure}[h]
   \centering
       \includegraphics[width=7cm]{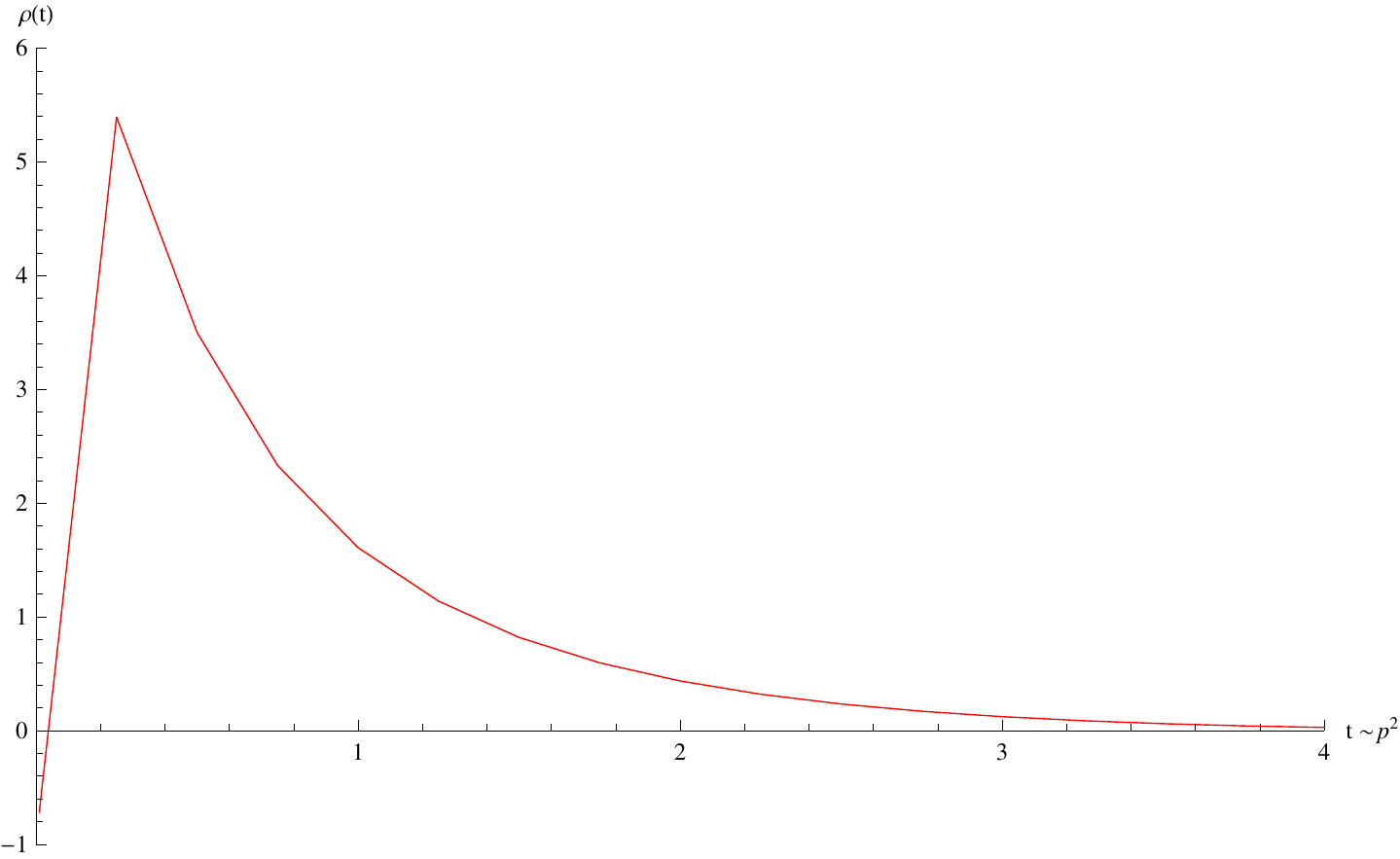}
               \caption{Spectral density for the (in principle non-existent) integral representation of the function (4.1).}
\end{figure}
Since the curve (\ref{1}) does describe the gluon lattice data very well and since the constructed ``mock gluon spectral density'' of Fig.~4 looks pretty different from the one presented in \cite{Strauss:2012dg}, this might suggest it would be favourable to further study the spectral behaviour of the gluon propagator. This is currently being undertaken using real lattice data input \cite{nieuw}.

\section{Conclusion}
We have briefly introduced an alternative approach to the problem of extracting spectral densities from numerical data and we tested it preliminarily on (1) the scalar glueball using gauge invariant lattice data and (2) fabricated gluon data in the Landau gauge. A more elaborate discussion and finer results will be presented elsewhere \cite{nieuw}.

\section*{Acknowledgments}
\noindent D.D.~is supported by the Research-Foundation Flanders (FWO Vlaanderen). P.~J.~S.~acknowledges support by FCT under contract SFRH/BPD/40998/2007. O.~O.~and P.~J.~S.~acknowledge financial support from the F.C.T. research project PTDC/FIS/ 100968/2008, developed under the initiative QREN financed by the UE/FEDER through the Programme COMPETE - Programa Operacional Factores de Competitividade. Part of the computation was performed on the HPC
clusters of the University of Coimbra. Part of the computational resources (Stevin Supercomputer
Infrastructure) and services used in this work were provided by Ghent University, the Hercules
Foundation and the Flemish Government - department EWI. We are grateful for technical support
from the ICT Department of Ghent University.

\end{document}